# Design of the photonic crystal demultiplexers for ultra-short optical pulses using the gap-maps analysis

I. A. SUKHOIVANOV[a*], I. V. GURYEV, O. V. SHULIKA, A. V. KUBLYK, O. V. MASHOSHINA, E. ALVARADO-MÉNDEZ[a], J. A. ANDRADE-LUCIO[a]

[a]FIMEE, University of Guanajuato, Salamanca, 36790, Mexico
Lab. "Photonics", National University of Radio Electronics, Kharkiv, 61166, Ukraine

In the work, the new method for the design of the wavelength division multiplexer on the basis of 2D photonic crystal integrated circuit for the ultra-short pulses channel separation is proposed and investigated. The method is based on the analysis of full photonic bandgap maps that allows defining the ultra-short pulse demultiplexer parameters selection. For the method approbation, there was synthesized the device in which the wideband filters are used for the channel separation. As it was expected, the device effectively separates 50 fs pulse channels with 1.31 and 1.55 microns wavelengths. The obtained spectral characteristics and pulse pattern responses validate the efficiency of the proposed method and the applicability of such devices to the integrated photonic information processing circuits.

## 1. Introduction

Today semiconductor electronics faced the challenge of the impossibility to improve the integrated devices performance that is connected with the amount of the logical elements size to its practical minimum and with increasing of electrical power requirements [1]. One can expect improving in this area if turn to development of principally new ways for the data handling, such as photonic crystal devices [2, 3].

Photonic crystals (PhCs), the structures with periodically changing refractive index (RI), can be used for creation of the alternative optical information processing devices. Due to their properties as well as structure peculiarities such as strict periodicity, the effect of strong light localization in the defect region of the structure appears [4]. The theoretical and the experimental investigations demonstrate that the PhCs can be applied to the wide-aperture single-mode lasers [5], wavelength filters[6], optical waveguide structures and sharp bends [7], wavelength division multiplexing (WDM) system devices [8, 9], splitters and combiners [10]. Thus, there is the possibility to build the full functional optical processor on the single PhC structure by the variation of the parameters of PhC and by the introducing the nonlinear elements [2]. Such optical devices promised to be compact, and can have high-performance.

One of the possible ways which is needed to be used to achieve the maximal performance of optical information processing device is the full utilization of the optical waveguide bandwidth. In order to maximize the information density in single physical channel the time-division multiplexing (TDM) and the WDM technique can be used in parallel [11]. To effectively realize the TDM it is convenient to use the ultra-short pulses (USP) technique [12]. However, USPs have very wide spectrum. For instance, the 100 fs USP with Gaussian shape has the spectrum width about 60 nm. The distance between channels in dense WDM systems is usually less than 0.2 nm [13]. Thus, it is impossible to use them for the USP channels separation.

As is known, wavelength filters of optical range based on one-, two- and three-dimensional photonic bandgap (PBG) structures [8], [14-16] can be created by the proper geometrical and physical parameters selection. As for optical frequency filters and moreover for demultiplexers, they can be based on the one-dimensional PhCs [9] in combination with optical circulators [17]. These demultiplexers consist of circulators placed one by one with precisely tuned Bragg reflectors between them. Such a technology allows to separate frequency channels with less than 100 GHz distance. The demultiplexer can also be realized using the high-Q nanocavities based on 2D PhCs [18]. However, none of such devices meet the requirements which are necessary for the USP demultiplexing. So the aim of the work consists in the development of the method that allows creating the devices on the basis of the 2D PhC which can provide wavelength demultiplexing of the pulse pattern with wide channels spectra.

The structure of paper is as follows. In the "Device design" section, the method for the design of the demultiplexer that uses wideband filters on the basis of the PhC as the wavelength filters was considered. Such devices allow demultiplexing the USP bit pattern. There was proposed the method for designing of such devices by the analysis of PBG maps. In the "Method examination" section, we synthesized the device for the separation of USP channels with wavelengths 1.31 and 1.55 μm using the proposed method. The obtained device characteristics validate the high-efficiency of the wavelength channel separation as well as the possibility of using such devices as elements of the integrated information processing optical devices.

## 2. Device design

The integrated optical device can be based on two types of 2D PhC. The first type uses the PhC structures having the background refractive index (RI) less than elements RI and the second one is vice-versa. In the first case the structure is "dielectric rods" (DR) type and in the second case is "perforated membranes". In both cases the degrees of freedom allowing the PBG managing are the geometry variations while the constant RI contrast and the RI contrast variation when the geometry remains constant. The combination of two mentioned ways as well as introducing the nonlinear optical effects can probably allow new possibilities. In most cases in equal conditions, the DR structures have larger PBG areas existing at technologically convenient geometric parameters when $r/a$ value is essentially lower than 0.5. Although the case of holes in a high index medium is still obviously more important in practice, one can meet difficulties with device realization using membranes. The reason is that in perforated slabs in most cases the PBG is formed at region of $r/a \sim 0.5$, that can result in problems when producing such structures. The exception is the PhC slabs with a triangular lattice. Moreover, as the additional computations have shown, perforated membranes have the PBGs much narrower than the dielectric rods at the same geometric parameters. In case of perforated membranes, the PBG achieves its maximum at the range of $r/a$ between 0.5 and 0.6. This actually corresponds to case of the DR but with the distorted rods shape. For instance, the case of circular holes with $r/a = 0.6$ is nothing else than the case of rods with the shape of concave square as it is shown in Fig. 1. Therefore, we restricted our analysis to the DR structures arranged in a square and triangular lattice. However, the method proposed here is general and does not introduce any special assumptions. So, it can be successfully applied to 2D PhC slabs as well.

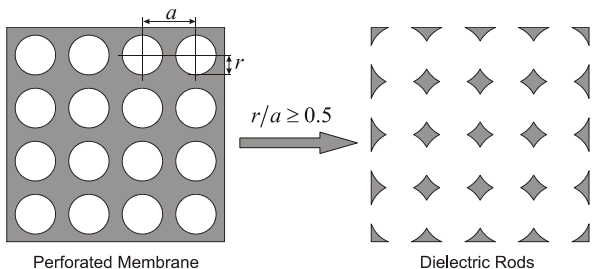

*Fig. 1. Identity of slab PhC to "dielectric rods" in case of $r/a > 0.5$.*

The two-channel demultiplexer considered here is T- or Y-splitter formed in W1 waveguide (1 defect line in the PhC). For the effective USP wavelength channels separation, the defects of special form in secondary waveguides are used. However, because of the wide USP spectrum, filters must have low quality and therefore wide transmittance spectrum. That's why the structures like add-drop multiplexers using microresonators and point defects [6,18] are inapplicable. In this case, it is necessary to use low-Q insertions which spectra do not override but lie inside the bandwidth of main channel.

The background PhC should form the bandgap as large as possible to support USP propagation in W1 waveguide. In order to obtain corresponding parameters value it is not enough just compute the PhC band structure. So the first design step is the computation of the PBG map, which can be obtained both by the variation of the geometry while the RI contrast remains constant and by the variation of the RI contrast while constant geometry. However, from the technological point of view, it could be easier to fabricate the structure with variation of geometry parameters than other one. In addition to this, while performing an exploratory analysis we found that geometry design is more suitable.

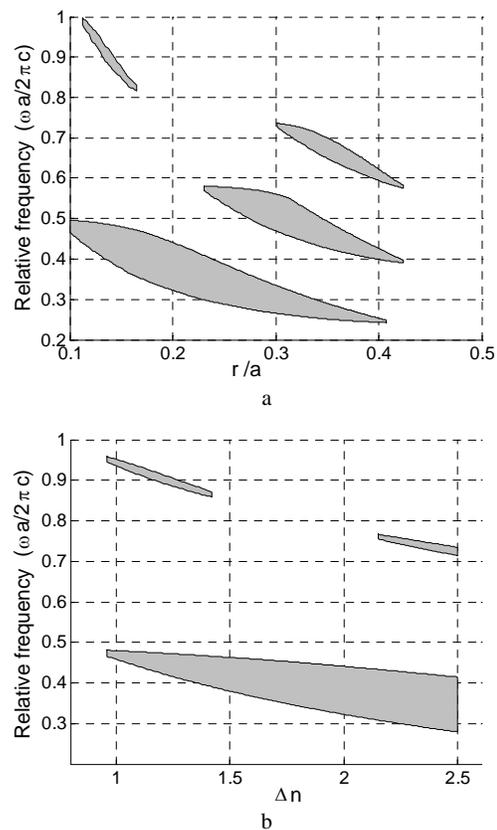

*Fig. 2. TM Bandgap maps for the square lattice PhC for different radius-to-pitch ratio (a), and for variation of the difference of refractive indices of rod and background materials (b).*

The results obtained using plane-wave expansion method [19] for the square lattice of rods with circular section embedded in air are shown in Fig. 2 we see that in case of geometry variation the central bandgap frequency can be changed in the range of 0.25 – 0.48 and in case of index contrast variation this range of 0.35 – 0.47. Besides, in case of index contrast variation the high-frequency edge of the bandgap changes much slower then the low-frequency one that results in nonvanishing PBG within the reasonable limits of index contrast which corresponds to the actual materials. Therefore, the contrast design is unsuitable for T- or Y-junctions. However, it may be applied to produce sequential filtering.

The obtained PBG map allows quick selection of the elements size and lattice constant of the 2D PhC which provides the largest bandgap for the structure. This corresponds to $r_2$ vertical line on Fig. 3. Then the edge frequencies $\Omega_1$ and $\Omega_2$ corresponding to bandgap edges of the background PhC can be found as cross-points of vertical straight line $r/a = r_2$ and PBG area edges, as shown in Fig. 3. Further we find the correspondence between the $\Omega_2$ and the short-wavelength part of transmittance spectrum of main. When we know the edge frequency $\Omega_2$, USP spectrum width, and the central USP wavelength, we are able to compute the lattice constant of the background PhC. The relative sizes of insertions in the secondary waveguides are defined by the cross-points of the horizontal line $\Omega(r/a) = \Omega_C$ and the PBG area edges, where $\Omega_C$ is the central relative frequency corresponding to the center of the primary waveguide bandwidth. We fix the lattice constant to be equal for both background PhC and defects. This could be convenient for fabrication process, but it is not requirement. The actual sizes of the insertions can be found via its relative sizes and lattice constant of the background PhC found at previous steps. Finally we should tune fine the defect sizes. We compute the transmittance at the corresponding wavelengths through the defects while radii of defects are varying in small range around the values obtained by the described method. The obtained radius with maximum transmission is utilized. We apply the method proposed for synthesis of the two-channel PhC demultiplexers for the USPs with central wavelengths of $\lambda_1 = 1.31 \mu m$ and $\lambda_2 = 1.55 \mu m$. The results are presented in the next section.

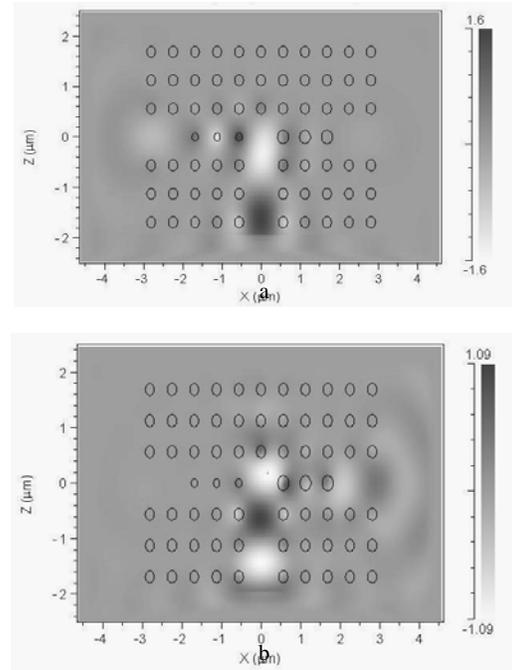

Fig. 4. Results of FDTD simulation of wavelength channel splitting: (a) – source wavelength $\lambda = 1.55 \mu m$, (b) - source wavelength $\lambda = 1.31 \mu m$.

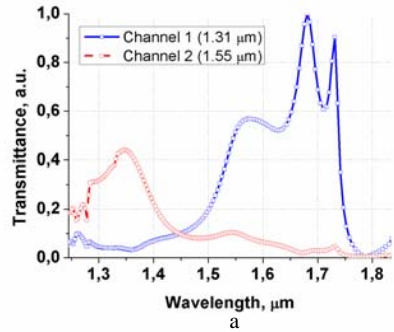

a

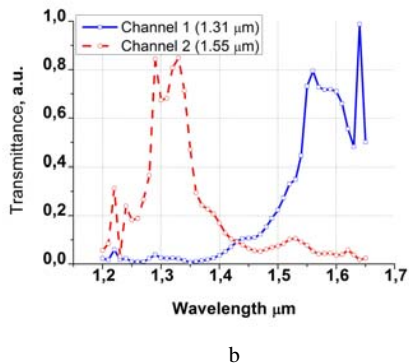

b

Fig. 5. Transmission spectra of the filters in secondary waveguides of the demultiplexer based on square lattice (a), and triangular lattice (b) PhCs.

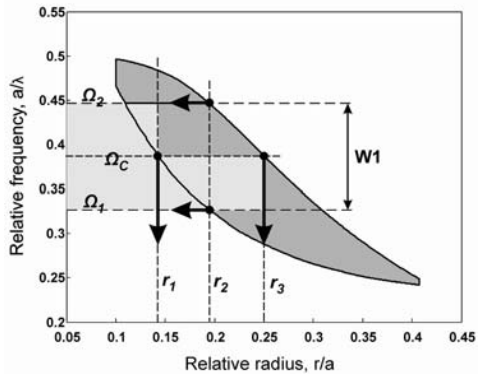

Fig. 3. The scheme for selection of geometric parameters using bandgap map

## 3. Method examination

Numerical simulations were made for structures with a square and triangular lattice. For the square lattice case, the background refractive index is $n_1 = 1$; the rod refractive index is $n_r = 3$. Following the method described we found for the square lattice case the lattice constant is $a = 0.563 \mu m$; the rods radii are $r_1 = 0.14a$; $r_2 = 0.2a$; $r_3 = 0.25a$. For triangular lattice case, background refractive index is $n_1 = 1$; the rod refractive index is $n_r = 3$; the lattice constant is $a = 0.65 \mu m$; the rods radii are $r_1 = 0.11a$; $r_2 = 0.17a$; $r_3 = 0.235a$. The computations were carried out using the FDTD method [20-22]. The field patterns are shown in Fig. 4 for the case of a square lattice structure. The filter in the left channel of the device fully reflects the radiation with 1.31 µm wavelength and the filter in the right channel – the radiation with 1.55 µm wavelength. To examine this precisely we computed transmittance spectra of secondary waveguides in square and triangular lattices, which are shown in Fig. 5. We can see that in case of square lattice the transmittance for the channel 1.55 µm is better than for 1.31 µm channel. For the case of a triangular lattice the PhC the situation is much better – the transmittance peaks for different wavelengths are almost identical.

We will illustrate fine-tuning by the example of square-lattice structure. The Fig. 6 shows the dependence of the transmittance on the rod radius at the wavelengths $\lambda = 1.55 \mu m$ and $\lambda = 1.31 \mu m$. These characteristics have good defined maxima. We took corresponding radii as final values to form incretions in secondary waveguides. Final values are $r_1 = 0.124a$, $r_2 = 0.275a$.

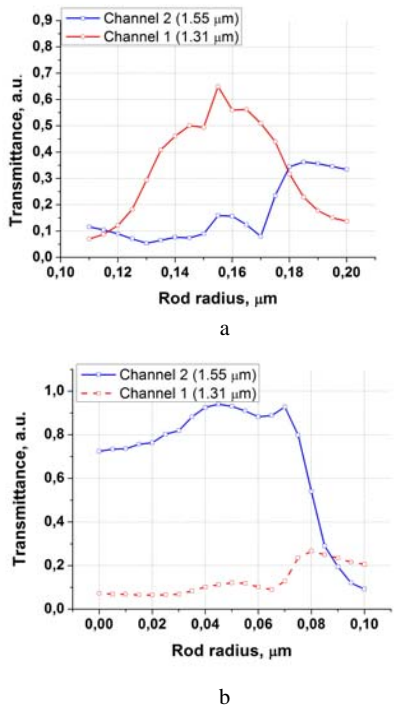

*Fig. 6. Tuning characteristics of the filters for $\lambda = 1.55 \mu m$ (a) and $\lambda = 1.31 \mu m$ (b). For comparison adjacent channels are shown.*

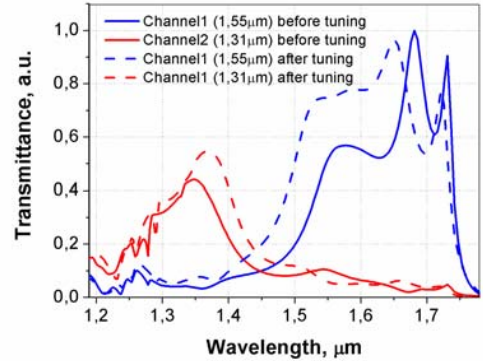

*Fig. 7. Transmittance spectra before and after the tuning. Square lattice*

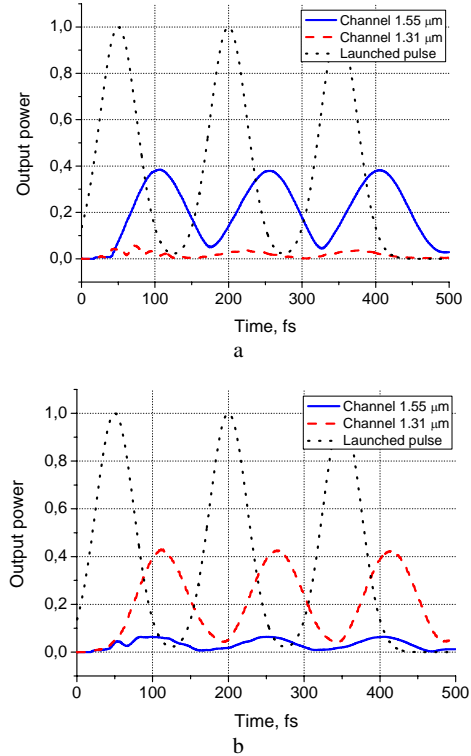

*Fig. 8. The pulse pattern responses of the demultiplexer based on the square lattice PhC. (a) – source wavelength $\lambda = 1.55 \mu m$; (b) – source wavelength $\lambda = 1.31 \mu m$. For comparison adjacent channels are shown.*

The Fig. 7 shows the comparison of initial T-structure to the fine-tuned one by transmittance spectra. As one can see tuning provide better spectra.

To estimate demultiplexer properties in the time domain we compute its response on pulse pattern at different wavelengths, which are shown in Fig. 8. It should be mentioned that the attenuation of the passed pulse equals approximately 60% in both cases of square and triangular lattice. But the broadening is almost absent, so the pulses remain clear enough to be recovered successfully. The pulse pattern response for the demultiplexer based on the triangular lattice PhC is similar to that of the square-lattice PhC.

The low broadening is achieved by applying wideband frequency filters. Though the employment of narrow-band filters provides multichannel demultiplexing, the temporal response of the device is expected to be poor due to the very broad spectrum of USP. Thus, we consider the proposed design conception as an effective one for the high-speed optical informational systems.

## 4. Conclusion

We proposed and examined simple method for design of the wavelength division demultiplexers on the basis of 2D PhC for USP processing. The method is based on the analysis of PBG map. The method proposed here does not introduce any special assumptions and, therefore, it can be applied to any 2D PhC structure. The synthesis of the structure has shown the efficiency of the proposed method. Spectral characteristics of the probe structure reveal the good wavelength separation for both types of lattices, square and triangular. FDTD simulations of the probe structure show effective device operation on separation of 50 fs optical pulses. Such results cannot be achieved by using the high-Q filters.

The results obtained in this work validate the possibility of using the method of the PBG maps analysis for the synthesis of the demultiplexers for the USP wavelength channel separation in the integrated PhC circuits.

---

[*]Corresponding author: i.sukhoivanov@ieee.org